\documentclass[aps,pra,a4paper,10pt,twocolumn,noshowpacs,superscriptaddress,floatfix]{revtex4-1}
\usepackage{graphicx}
\usepackage{amsmath}
\usepackage{amsfonts}
\usepackage{amssymb}
\usepackage[colorlinks,
            linkcolor=blue,  
	     anchorcolor=blue,
	     urlcolor=blue,
	     citecolor=blue]{hyperref}

\makeatletter
\renewcommand{\maketag@@@}[1]{\hbox{\m@th\normalsize\normalfont#1}}%
\makeatother\usepackage{braket}

\begin{document}

\title{Collective resonance in helical superstructures of gold nanorods}
\author{Xuxing Lu}
\thanks{Current Address: Huygens-Kamerlingh Onnes Laboratory, Leiden University, 2300 RA Leiden, Netherlands}
\email{lu@physics.leidenuniv.nl}
\author{Weixiang Ye}
\email{wxye@suda.edu.cn}
\author{Wenlong You}
\author{Hao Xie}
\author{Zhihong Hang}
\author{Yun Lai}
\author{Weihai Ni}
\email{niweihai@suda.edu.cn}

\address{Jiangsu Key Laboratory of Thin Films, School of Physical Science and Technology, Soochow University, Suzhou, Jiangsu, 215006, China}

\begin{abstract}
    Chiroptical responses of helical superstructures are determined by collective behaviors of the individual building blocks. In this paper, we present a full theoretical description of the collective resonance in superstructures. We use the gold nanorods as individual building blocks and arrange them helically along an axis in an end-to-end fashion.  Numerical simulations on single-unit cells reveal that the plasmonic coupling between the nanorods produces hybridized resonances, whose intensity is strongly dependent on the excitation light with left- or right-handed circular polarizations (LCP or RCP). A node-mode criterion is proposed on the basis of the microscopic mechanism, which successfully explains the difference between LCP and RCP. We further demonstrate, by repeating the unit cell from 1 to infinity along the helical axis, the multiple hybridized resonances gradually evolve and merge into a single collective resonance, whose energy is also dependent on LCP and RCP. An analytical description is provided for the collective resonance of the helical superstructure on the basis of the coupled dipole approximation method. Our theory shows that n collective resonance modes are present in the helical superstructure with the unit cell consisting of $n$ nanorods. Strikingly, only one resonance can be excited by the incident light with certain circular polarization. We propose a universal selection rule for such selective excitation of the collective resonances by analyzing the symmetry of the helical superstructures. The new insights provided in this work may shed light on future designs and fabrications of helical superstructures using plasmonic building blocks.
\end{abstract}
\maketitle
\section{introduction}
Chirality refers to a symmetric property possessed by a category of objects that cannot be superimposed with their mirror images like our hands. These objects include amino acids,proteins, sugar molecules, and etc. Chirality is able to manifest itself optically by responding differently to incident light with left- or right-handed circular polarizations (LCP or RCP), giving rise to differential absorption called circular dichrism (CD)\cite{Hentschele1602735}. CD spectroscopy serves as a powerful tool in the field of chemistry, biochemistry, biology, and physics for probing the stereoscopic information of the chiral objects, such as secondary structure and conformation of macromolecules\cite{barron2009molecular,amabilino2009chirality}. However, the CD responses of natural molecules are relatively weak due to the small electromagnetic volume, which handicaps its further applications.

For a chiral molecule, the chirality stems from a time-even pseudoscalar, the rotational strength, which can be represented by the imaginary part of the dot product of an electric dipole $\mathbf{p}$ (time-even vector) and a magnetic dipole $\mathbf{m}$ (time odd pseudovector)\cite{schellman1975circular,luo2017plasmonic}. The enhancement of chiral response by the local electric field $\mathbf{E}$ and magnetic field $\mathbf{B}$, is proportional to a local parameter $C$, the optical chirality (OC) of the electromagnetic field\cite{lipkin1964existence,tang2010optical}. Therefore, the CD responses of the small molecules can be greatly enhanced by generating superchiral electromagnetic fields\cite{tang2011enhanced}.  Recently, plasmonic analogs of chiral molecules have stimulated significant scientific interests for understanding and enhancing the CD responses of the small molecules\cite{liu2014helical,guerrero2012molecular,lu2013discrete,smith2016chiral,wang2017circular,wu2013unexpected,zhao2017chirality}. The main reason is the extremely large dipole strength of plasmonic nanoparticles that are associated with plasmonic oscillations of the quasi-free conduction electrons. While the electronic or vibrionic excitations of the secondary structures account for the chiroptical responses in natural molecules, plasmonic oscillations of conduction band electrons endow unique chiroptical responses to metallic chiral nanostructures with flexibility in varying their shapes, sizes, configurations, or compositions. They are promising in the applications of enhancing the CD responses of molecules nearby\cite{schaferling2012tailoring,schaferling2014helical,hendry2010ultrasensitive}, as well as novel circular polarization-selection optical devices\cite{gansel2009gold,zhao2012twisted,yin2015active}. Substantial efforts have been devoted to the designs and fabrications of three-dimensional (3D) plasmonic structures, which include both top-down\cite{hentschel2012three,yin2013interpreting,cui2014giant,hentschel2015optical,deng2018helical,ryu2019helical,frank2013large,ferry2015circular} and bottom-up approaches\cite{acuna2012fluorescence,acuna2012distance,puchkova2015dna,pellegrotti2014controlled,urban2018gold,mokashipunekar2019deliberate,kuzyk2012dna,shen2011rolling,cheng2017goldhelix,merg2016peptide-directed,song2013tailorable,urban2016plasmonic,shen2013three,dai2014dna,yan2012self,lan2013bifacial,kuzyk2016light,liu2019colloidal,funck2018sensing,huang2018dna,urban2015optically,guerrero2011intense,lan2017programmable,lan2018dna-guided,funck2018sensing,lan2014nanorod,wang2013giant,shen2015tuning, liu2019colloidal,huang2018dna,kuzyk2018dna,hu2019plasmonic}. Examples range from chiral nanostructures\cite{gansel2009gold,frank2013large} to the chiral arrangement of a achiral building blocks such as gold nanospheres\cite{kuzyk2012dna,dai2014dna,shen2013three,shen2011rolling,cheng2017goldhelix,merg2016peptide-directed,yan2012self,song2013tailorable,urban2016plasmonic}, nanorods\cite{lan2013bifacial,kuzyk2016light,urban2015optically,lan2014nanorod,shen2015tuning,yin2015active,guerrero2011intense,lan2017programmable,lan2018dna-guided,liu2019colloidal,funck2018sensing,wang2013giant,huang2018dna} and both\cite{ferry2015circular,mokashipunekar2019deliberate}. It is worth noting that pronounced CD responses can also be observed in two-dimensional (2D) achiral structures excited normally\cite{zu2016planar,schnell2015real} or obliquely\cite{plum2009metamaterials,gompf2011periodic,lu2014circular} by the incident CPL. In principle, a common essential feature of all the aforementioned configurations is the geometric chiral variations on the propagation direction of the incident circularly polarized light (CPL).

\begin{figure}[tpb]
    \centering
    \includegraphics[width=0.9\linewidth]{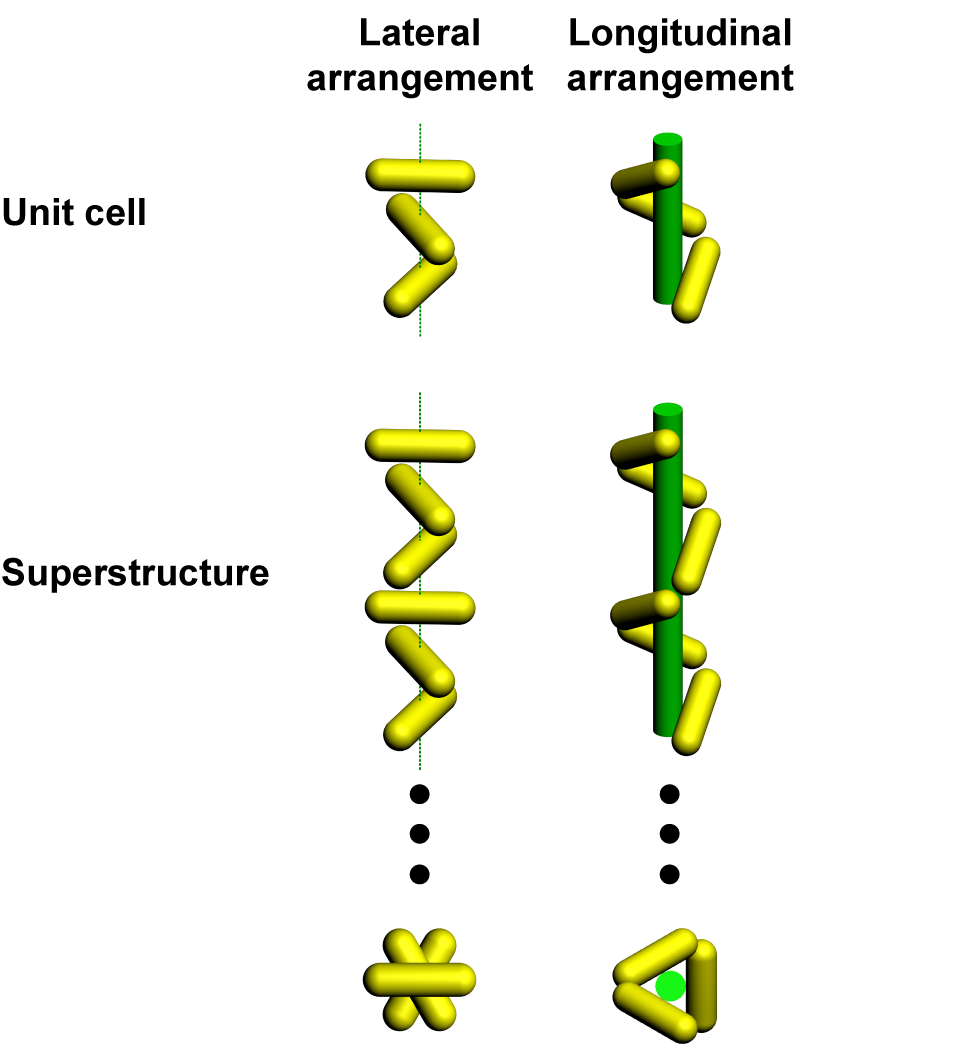}
    \caption{Schematic illustration of gold nanorod helical superstructures with lateral and longitudinal arrangements. The unit cell consists of two side-by-side or end-to-end assembled nanorods, which exhibiting chiroptical response under the excitation of CPL. The unit cell consisting of n nanorods is repeated in the z-direction to form a helical superstructure. The bottom of the figure shows that the $n$ nanorods form a circle with an in-plane angle $\varphi_n^0=\frac{2\pi}{n}$  between adjacent nanorods when projected to the x-y plane.}
    \label{fig:helix_mode}
\end{figure}

One interesting feature of the plasmonic analogs of chiral molecules is their collective resonance, which is the sum effect contributed by individuals in an array or a superstructure. The collective resonance has been considered as an important factor for generating sharp spectral features in the optical response\cite{auguie2008collective}. While the geometric chiral configuration of individual structures has a remarkable impact on their chiroptical response, the overall chiroptical response of a helical superstructure is actually determined by the collective behavior of the individuals, which is dependent on their distance and mutual orientations. Individual features will diminish and give way to the collective ones when the individual unit is repeated to form a superstructure. Some helical structures with multiple pitches have been successfully prepared experimentally\cite{gansel2009gold,kuzyk2012dna,shen2011rolling,lan2014nanorod,guerrero2011intense,song2013tailorable,urban2016plasmonic,wang2013giant,cheng2017goldhelix,merg2016peptide-directed,mokashipunekar2019deliberate,lan2017programmable,lan2018dna-guided}. Unfortunately, only a few reports concern the theoretical aspects of the collective resonance in the plasmonic helical superstructures\cite{christofi2011spiral-staircase,christofi2012giant,wang2015plasmonic,tserkezis2018circular}. In recent years, gold nanorods have been chosen as the building blocks for plasmonic helical superstructures.  They are ideal candidates for a model system with chirality because they exhibit the longitudinal plasmonic mode featured by dipolar oscillation specifically along their long axis\cite{lan2013bifacial,kuzyk2016light,urban2015optically,lan2014nanorod,shen2015tuning,yin2015active,guerrero2011intense,lan2017programmable,lan2018dna-guided,liu2019colloidal,funck2018sensing,wang2013giant,huang2018dna}. Gold nanorod helical superstructures can be classified into weak and strong coupling cases, respectively corresponding to the lateral and longitudinal arrangements (FIG.~\ref{fig:helix_mode}). The chiroptical responses from the laterally\cite{auguie2011fingers,smith2016chiral} and longitudinally\cite{lu2014circular} arranged nanorod dimers have been analyzed in detail. Recently, we have provided a straightforward explanation for the chiroptical response from the helical superstructure of the lateral arrangement\cite{lan2014nanorod}. However, a simple model to describe the chiroptical responses of the superstructures is still in vain.

Herein, we provide a detailed theoretical description for collective resonance in typical helical gold nanorod superstructures. We consider the strong coupling case where adjacent nanorods are longitudinally arranged in an end-to-end fashion. The strong coupling is interesting and worthy of investigation because its chiroptical response is much stronger than the weak coupling case. The nanorods in the unit cell are helically arranged in an end-to-end fashion and form a circle when projected to the x-y plane (FIG.~\ref{fig:helix_mode}). The unit cell is repeated in the z-direction to form a helical superstructure. Numerical simulation was firstly started with single-unit cells formed by three and four nanorods by employing a finite element method (Comsol Multiphysics). To explain the difference of  the chiroptical responses between LCP and RCP, we proposed a simple node-mode criterion on the basis of the microscopic mechanism. The simulation was further continued with a superstructure formed by repeating the unit cell along the helical axis. The evolution of the chiroptical response was investigated when the number of unit cells was gradually increased from 1 to infinity. We provided an analytical description for the collective resonance mode of the infinite helical structure on the basis of the coupled dipole approximation (CDA) method (see APPENDIX), and a selection rule for the selective excitation of the collective resonances by analyzing the symmetry of the helical superstructure. Compared to other theoretical models based on CDA\cite{christofi2012giant,tserkezis2018circular} or based on full-electrodynamic layer-multiple-scattering (LMS) method \cite{christofi2011spiral-staircase}, our model is much straightforward in investigating light interaction with helix structures, as we can reduce the degree of freedom by fixing the dipole orientation of each nanorod along their longitudinal axis. Our model allows one to predict the resonance energy of the collective mode of a well-defined superstructure and their optical response qualitatively without calculating the whole spectrum. Besides, this model can be generalized to more general cases by simply changing the states of incident light or the orientations and locations of the gold nanorods. Our new insights may open up a new way not only for future designs and fabrications of helical superstructures using plasmonic building blocks, but also for the study of chirality interaction in other low dimensionnal systems.

\section{Numerical simulation}
We investigate the strong coupling case in the longitudinal arrangement as an example in this work ( the right panel in FIG.~\ref{fig:helix_mode}). The unit cell of the superstructure is formed by n discrete nanorods, and for simplicity, n is set to be an integer larger than 2. The nanorods circles around the axis of the helix (z-axis) with an orbital radius of and displacement of $D_z/n$ on the z-direction, where $D_z$ is the helical pitch. The $n$ nanorods form a circle with an in-plane angle $\varphi_n^0=\frac{2\pi}{n}$  between adjacent nanorods when projected to the x-y plane. Therefore, the primitive cell must be a complete circle of a certain number of nanorods when projected to the xy plane. They are tilted out of the x-y plane at a helix angle of $\theta$ so that the nanorods are arranged longitudinally. It requires the pitch length is determined jointly by the length of the nanorods, the helix angle, and the gap distance between adjacent nanorods. The superstructure is formed when the unit cell is repeated and sequentially placed along the z-axis. A valid helical superstructure of the longitudinally arranged nanorods requires that the orbital radius  and the pitch of the helix $D_z$ must obey the following relations
\begin{equation}
R = \frac{1}{2}\tan\frac{\beta}{2}\left[\frac{2r+\delta}{\sqrt{\tan^2\theta+\sin^2 \frac{\beta}{2}}}+l\cos\theta\right],
    \label{R}
\end{equation}
and
\begin{equation} 
    D_z/n = l\sin\theta + \left[\frac{(2r+\delta)\tan\theta}{\sqrt{\tan^2\theta+\sin^2(\frac{\beta}{2})}}\right],
    \label{Dz}
\end{equation}
where $L$ and $r$ represent the length and radius of the nanorod respectively, $\beta=\pi-\varphi_n^0$, $l=L-2r$, and $\delta$ is the gap distance between adjacent nanorods.

\begin{figure}[tbp]
    \flushleft
    \includegraphics[width=1.0\linewidth]{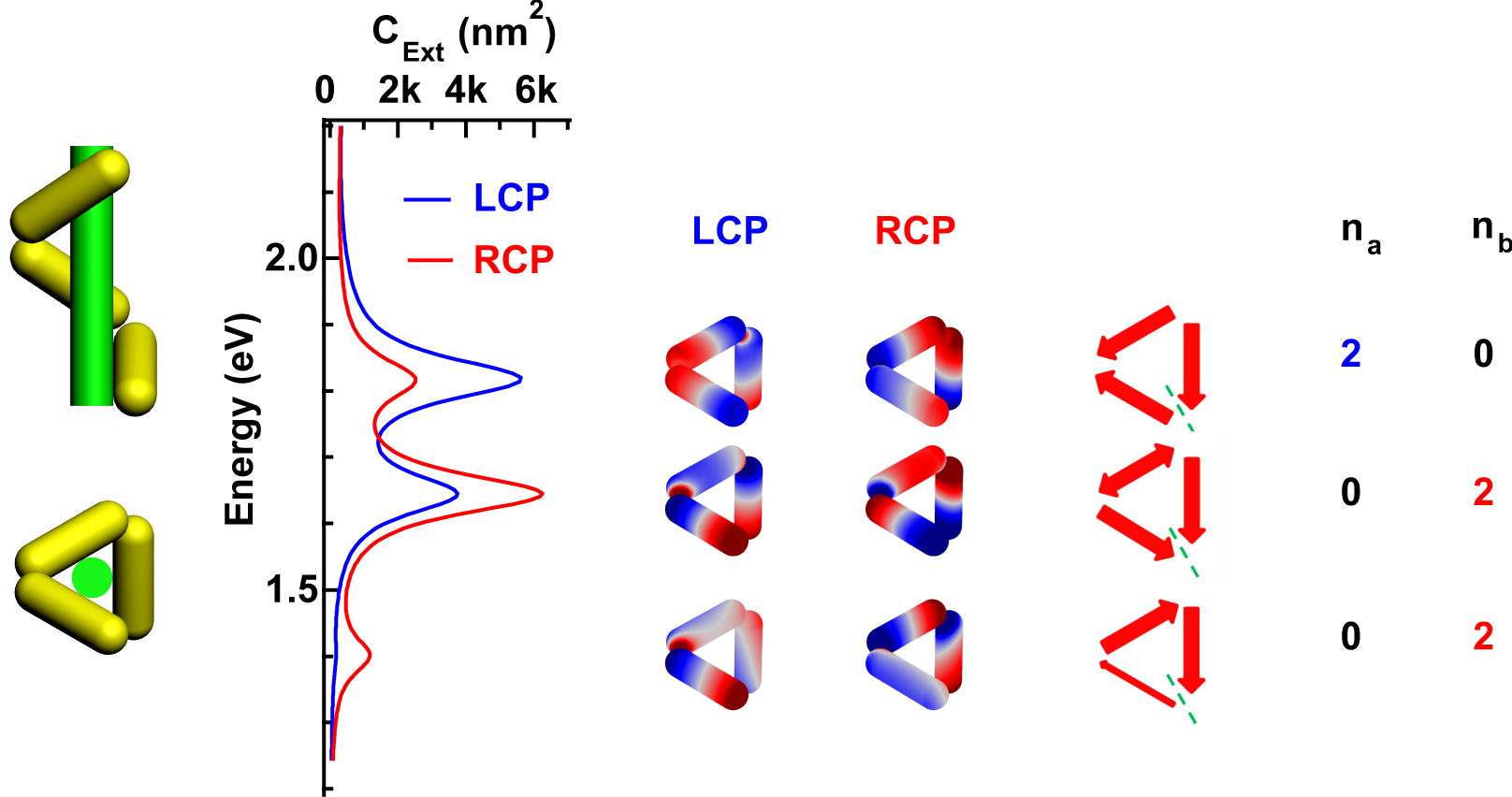}
    \caption{FEM-simulated chiroptical responses of the single-unit cell consisting of three nanorods. The extinction spectra of the single-unit cell excited by LCP (blue line) and RCP (red line) exhibit three hybridized resonance peaks located at $1.40$, $1.64$, and $1.82~ \mathrm{eV}$, respectively. The middle panel of the figure demonstrates the charge distribution profiles of the helical structure at the three hybridized energies with the LCP and RCP excitations. The number of anti-bonding and bonding nodes, na and nb, are counted and compared at the three hybridized resonance energies. The green dotted line on the left panel of the figure indicates the opening position of a single spiral structure, and the thickness of the arrow indicates the strength of the oscillation mode of the dipole. In our simulations, the diameter and length of the gold nanorods were set at 12 and 14 nm respectively. The refractive index of the ambient medium was set at 1.33. The gap distance between adjacent nanorods is 1 nm, and the helix angle is $30^{\circ}$.}
    \label{fig:helix_3_mode}
\end{figure}

We begin our simulation with the simplest single-unit cell formed by three nanorods with a right-handed arrangement is considered. The finite element method (FEM, Comsol Multiphysics) was employed to investigate the chiroptical response of the helical structure. The nanorod was modeled as a cylinder with two hemispheres at both ends, with the diameter $r$ and length $L$ set as $6$ and $40~\mathrm{nm}$ respectively. The nanorods are surrounded by a homogeneous medium (in our paper, the medium is water with a refractive index of $1.33$). The gap distance  between adjacent nanorods is $1~\mathrm{nm}$, and the helix angle  is $30^{\circ}$. The $D_z$ and $R$ will be $122.20$ ($141.55$) and $13.45$ $(20.62)~\mathrm{nm}$ for the superstructure is formed by $3~(4)$ discrete nanorods. For sake of simplicity, we only consider the situation that the CPL propagates along the direction of the helical axis. FIG.~\ref{fig:helix_3_mode} shows the FEM-simulated extinction spectra of the single-unit cell under the excitation of incident light with LCP and RCP. It can be seen that three resonance peaks locate at $1.40$, $1.64$, and $1.82~\mathrm{eV}$ are present in the extinction spectra due to the hybridization of plasmonic resonances of the gold nanorods\cite{nordlander2004plasmon}. Please noted that their intensities of those three resonance peaks are remarkably different under LCP and RCP excitations. The LCP resonance peak at the highest energy of $1.82~\mathrm{eV}$ exceeds that of the RCP, while at the other two lower energies of $1.64$ and $1.40~\mathrm{eV}$ the RCP response is stronger. To elucidate this polarization dependence of the resonance, the charge distribution profiles of the helical structure are plotted at the three resonance energies for the LCP and RCP excitations. As shown in FIG.~\ref{fig:helix_3_mode}, the LCP and RCP excitations do not have a significant difference in charge oscillation modes. However, the oscillations at different energies are remarkably distinctive. Analysis of the nodes formed by two adjacent nanorods will help us to get a deep insights into the oscillation modes as well as their corresponding chiroptical response under certain excitations\cite{lu2014circular}. Each node can deemed as a capacitor separated by a small gap where the charge distributions of proximal ends are either the same or opposite in sign, which is defined as the anti-bonding or bonding modes of the node in a similar way as the V-shaped nanorod dimers\cite{lu2014circular}. In this way, oscillating modes can be readily identified and analyzed in a complicated helical structure comprising multiple nodes. For example, two nodes can be identified for the helical structure investigated here, and the node-modes can be either bonding or anti-bonding. In our paper,  $n_\mathrm{a}$ and $n_\mathrm{b}$ represent the number of anti-bonding and bonding nodes, respectively. At the highest hybridized energy $1.82~\mathrm{eV}$, all the three nanorods behave as electric dipoles with the charge oscillation along their long axes, and adjacent nanorods oscillate in the opposite phase corresponding to anti-bonding mode at each node (FIG.~\ref{fig:helix_3_mode}). Therefore two anti-bonding nodes are present and $n_\mathrm{b} = 2$ (FIG.~\ref{fig:helix_3_mode}), which results in the highest resonance energy. At the moderate energy 1.64 eV, two rods at the ends of the helix oscillate as electric dipoles in the opposite phase, which induces a quadruple oscillation in the middle rod. Thus this scenario can be seen as two bonding nodes oscillating in the opposite phase, and $n_\mathrm{b} = 2$, which results in lower resonance energy. At the lowest energy $1.40~\mathrm{eV}$, the chiroptical response and the oscillation modes strongly depend on the polarization. Under the excitation of RCP, strong and weak bonding nodes in the same phase are present, resulting in a net sum effect of the bonding mode as well as the lowest resonance energy. As for LCP, the mode contains a dipole asymmetrically coupled to a weak bonding node, which is so inefficient that the resonance peak is hardly observable. 

\begin{figure}[tpb]
    \centering
    \includegraphics[width=1.0\linewidth]{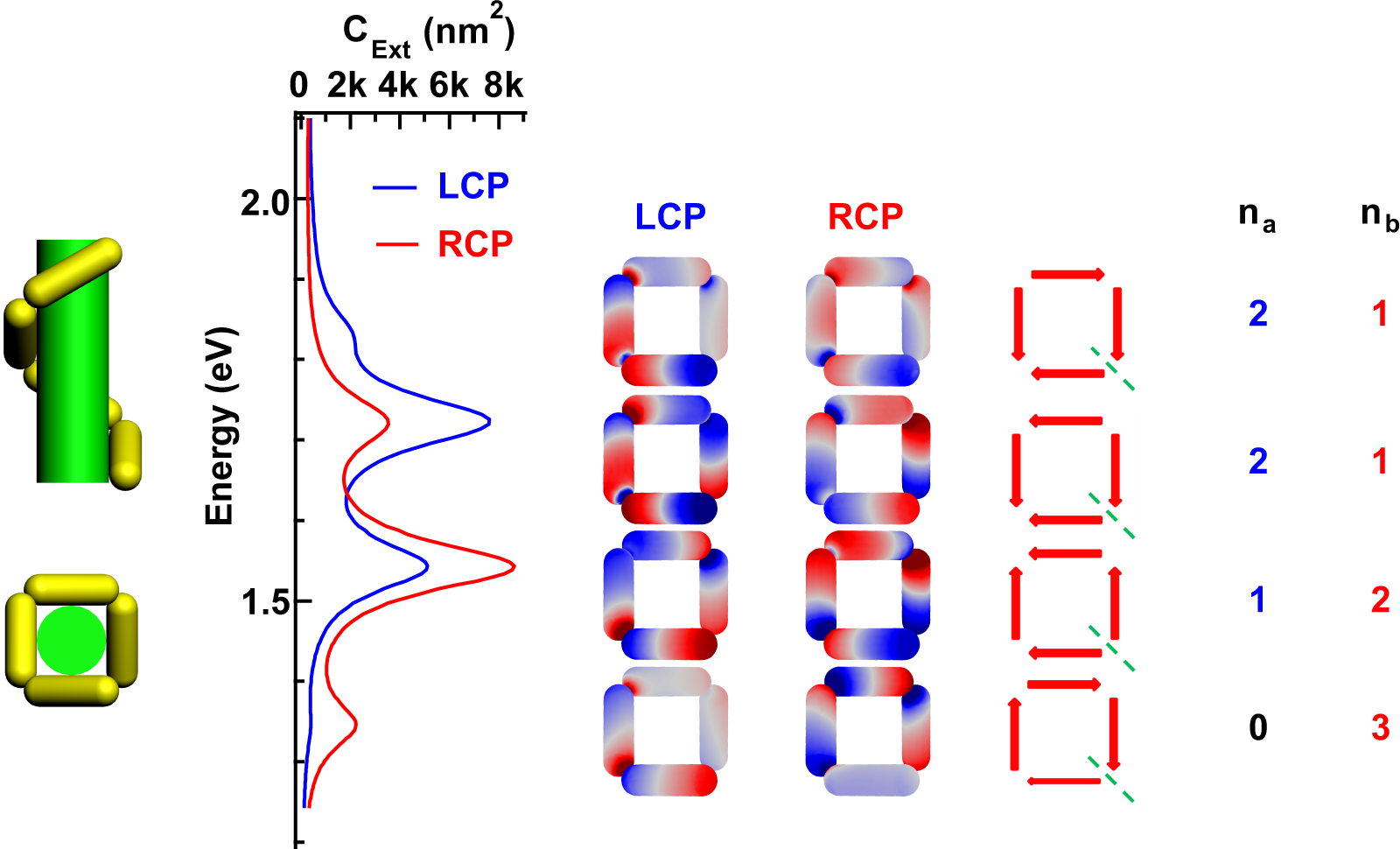}
    \caption{FEM-simulated chiroptical responses of the single-unit cell consisting of four nanorods. The extinction spectra of the single-unit cell excited by LCP (blue line) and RCP (red line) exhibit four hybridized resonance peaks located at $1.34$, $1.54$, $1.72$, and $1.82~ \mathrm{eV}$ respectively. The charge distribution profiles of the helical structure are plotted at the four resonance energies for the LCP and RCP excitations. The number of anti-bonding and bonding nodes, $n_\mathrm{a}$ and $n_\mathrm{n}$, are counted and compared at the four resonance energies.}
    \label{fig:helix_4_mode}
\end{figure}

We further consider a single-unit cell comprising four nanorods with the right-handed arrangement. Simulation parameters of the helical structure including the gap distance and helix angle are the same as the case of three nanorods. FIG.~\ref{fig:helix_4_mode} shows the FEM-simulated extinction spectra under the excitation of incident light with LCP and RCP, where four resonance peaks are excited respectively at $1.34$, $1.54$, $1.72$, and $1.82~\mathrm{eV}$. It can be noticed that the two hybridized resonances at the highest ($1.82~\mathrm{eV}$) and the lowest ($1.34~\mathrm{eV}$) energies response only to LCP and RCP respectively, while the other two resonances at the energies of $1.72$ and $1.54~\mathrm{eV}$ response to both and are in favor of LCP and RCP respectively. The charge distribution profiles (FIG.~\ref{fig:helix_4_mode}) illustrate that most of the nanorods behave as dipoles which oscillate in different modes corresponding to the resonance energies. In order to characterize these modes, we perform detail analysis of the aforementioned node. Three nodes are present in the structure of four nanorods. At the highest energy of $1.82~\mathrm{eV}$, the nodes are respectively bonding, anti-bonding, and anti-bonding from bottom to top and therefore $(n_\mathrm{a}, n_\mathrm{b}) = (2, 1)$. At $1.72~\mathrm{eV}$, two anti-bonding nodes are separated by a bonding node. Although the values of $n_\mathrm{a}$ and $n_\mathrm{b}$ are the same as those at $1.82~\mathrm{eV}$, the resonance energy is lower because the bonding node in the middle provides a channel for the two anti-bonding nodes to communicate. In contrast, at $1.82~\mathrm{eV}$ the two anti-bonding nodes are interconnected to each other accounting for its highest resonance energy. At $1.54~\mathrm{eV}$, $(n_\mathrm{a}, n_\mathrm{b}) = (1, 2)$, and therefore the resonance energy is further lowered. At $1.34~\mathrm{eV}$, only three bonding nodes are identified resulting in the lowest resonance energy.

A close investigates at the spectra and the charge oscillation modes lead to the direct correlation between the polarization-dependent excitation efficiency and the node oscillation modes. Our previous work reveals that in case of right-handed configuration (formed together by the incident beam and the dimer), the RCP excitation efficiency of the bonding mode is higher than the LCP, and that of the LCP is higher for the anti-bonding mode\cite{lu2014circular}. Therefore, in cases of $n_\mathrm{a}>n_\mathrm{b}$, the LCP excitation efficiencies must always be higher than the RCP, where the relation of $C_\mathrm{ext,L}>C_\mathrm{ext,R}$ is valid, and vice versa. On the basis of this simple criterion, one can readily judge the excitation efficiency of LCP and RCP at a certain resonance frequency by counting and comparing the numbers of $n_\mathrm{a}$ and $n_\mathrm{b}$. For example, as displayed in FIG.~\ref{fig:helix_4_mode}, at $1.72$ and $1.82~\mathrm{eV}$, $(n_\mathrm{a}, n_\mathrm{b}) = (2,1)$, and the response of LCP is much stronger than the RCP, while at $1.34$ and $1.54~\mathrm{eV}$, $(n_\mathrm{a}, n_\mathrm{b}) = (1,2)$ and $(na, nb) = (0,3)$, the response of RCP is stronger. \textbf{This criterion is universal and works well for other helical structures.} As illustrated in FIG.~\ref{fig:helix_3_mode}, at $1.82~\mathrm{eV}$, $(n_\mathrm{a}, n_\mathrm{b})= (2,0)$, and $C_\mathrm{ext,L}>C_\mathrm{ext,R}$ , while at $1.40$ and $1.64~\mathrm{eV}$, $(n_\mathrm{a}, n_\mathrm{b}) = (0,2)$, $C_\mathrm{ext,L}<C_\mathrm{ext,R}$. Although at $1.64~\mathrm{eV}$ quadruple mode is excited in one of the nanorods, the criterion is still valid in determining the excitation efficiency.

We propose a microscopic mechanism for describing the node-mode criterion on the basis of the understanding of the chiroptical response from the V-shaped nanorod dimer under oblique excitation\cite{lu2014circular}. The helical structure investigated here can be viewed as a stack of V-shaped nanorod dimers oriented and tilted in sequence at varying angles. The V-shaped nanorod dimer, when tilted to the incident CPL, exhibits distinctive chiroptical responses at hybridized bonding and anti-bonding resonance energies depending on the symmetry of the hybridized oscillation and its relative orientation to the incident light. In the view of the incident CPL, the two nanorods are placed one after another with their dipole moments parallel or antiparallel to the spatial evolution of the electric field vectors of the CPL, resulting in a distinctive chiroptical response. This picture is analogous with the Born-Kuhn model for the L-shaped nanorod dimer with vertical displacement\cite{yin2013interpreting}. In the case investigated here, the dipole moments of the anti-bonding modes are favorable of LCP, while on the contrary, those of the bonding modes are favorable of RCP. The helical structure comprises a number of nodes on either bonding or anti-bonding modes with each node contributing equally to the chiroptical response of the structure, and therefore the evaluating of the response at a certain polarization state can be performed by counting and comparing the numbers of the bonding and anti-bonding nodes.

\begin{figure}[tpb] 
    \centering
    \includegraphics[width=1.0\linewidth]{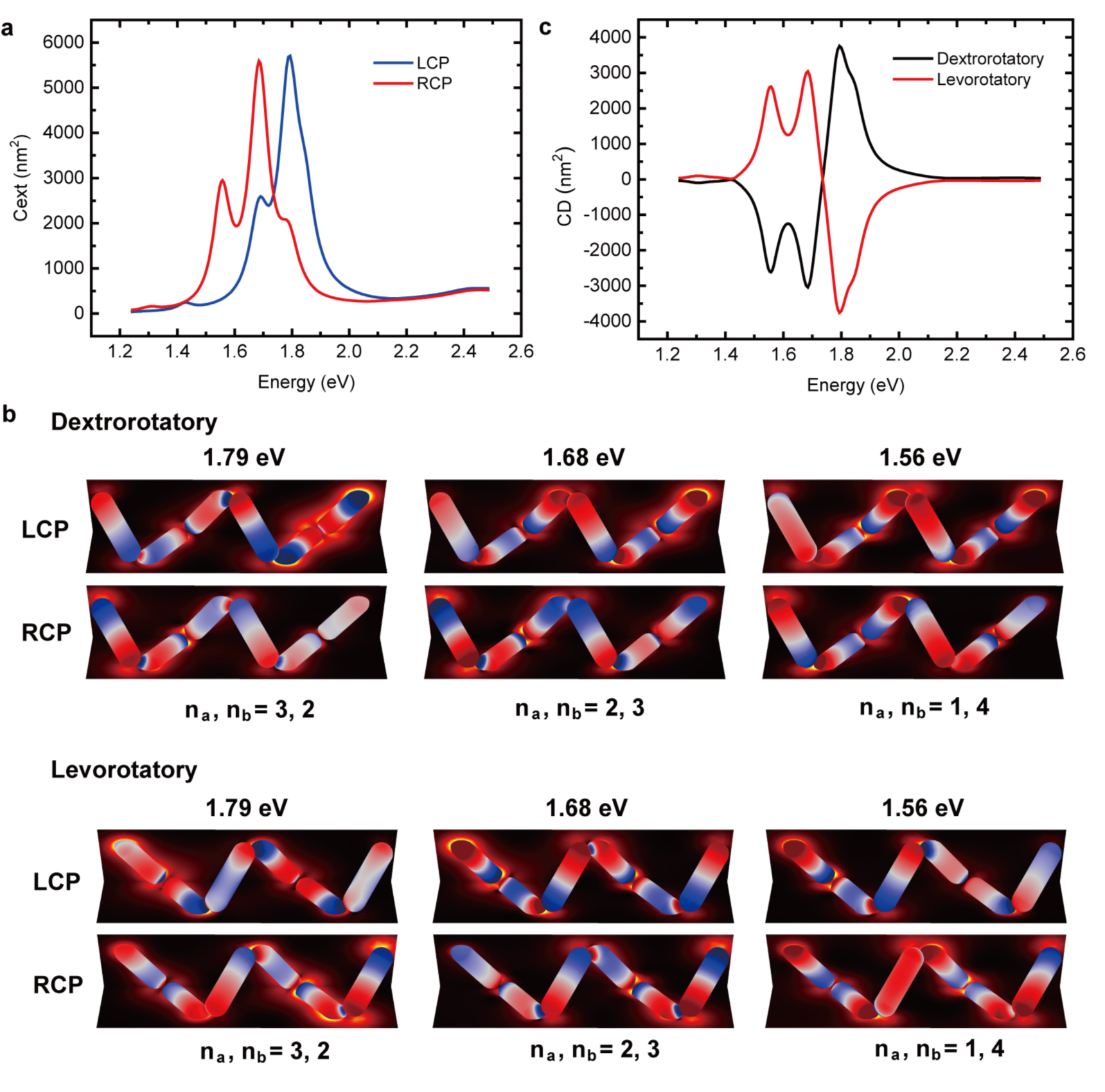}
    \caption{The chiroptical responses of two successive unit cells ( the single-unit cell consisting of three nanorods) from FEM-simulation. (a) The extinction spectra of the helical structure under the excitation of incident light with LCP (blue line) and RCP (red line) shows six hybridized peaks are located at $1.31$, $1.43$, $1.56$, $1.68$, $1.79$, and $1.84~\mathrm{eV}$, respectively. (b) LCP and RCP charge distribution profiles of dextrorotatory and levorotatory nanostructures at $1.56$, $1.68$, and $1.79~\mathrm{eV}$. The numbers of nodes $(n_\mathrm{a},n_\mathrm{b})$ at these energies are (3,2), (2,3), and (1,4), respectively. (c) CD spectra of the dextrorotatory and levorotatory helical structures.}
    \label{fig:Fig4_helix3_2}
\end{figure}

To further verify the above criterion, we continue with a helical structure consisting of two successive unit cells. The unit cell is formed by three nanorods as described before. Six hybridized peaks are excited respectively at $1.31$, $1.43$, $1.56$, $1.68$, $1.79$, and $1.84~\mathrm{eV}$ in the extinction spectra (FIG.~\ref{fig:Fig4_helix3_2}a). Among the hybridized peaks, the peaks located at $1.56$, $1.68$, and $1.79~\mathrm{eV}$ are significantly stronger, and hence the charge distribution profiles of LCP and RCP at these energies are depicted in FIG.~\ref{fig:Fig4_helix3_2}b. The numbers of nodes $(n_\mathrm{a},n_\mathrm{b})$ at these energies are $(3,2)$, $(2,3)$, and $(1,4)$, respectively, which suggests LCP dominating at $1.56~\mathrm{eV}$, and RCP dominating at $1.68$ and $1.79~\mathrm{eV}$. This prediction based on the criterion again agrees well with the relative intensities of the LCP and RCP peaks in the spectra. Moreover, we also consider the case of the left-handed (levorotatory) helix structure. With the same status of the nodes, a vertically mirrored CD signal is obtained (FIG.~\ref{fig:Fig4_helix3_2}c) indicating the opposite form of the criterion in the left-handed configuration.

\begin{figure}[tpb]  
    \centering
    \includegraphics[width=1.0\linewidth]{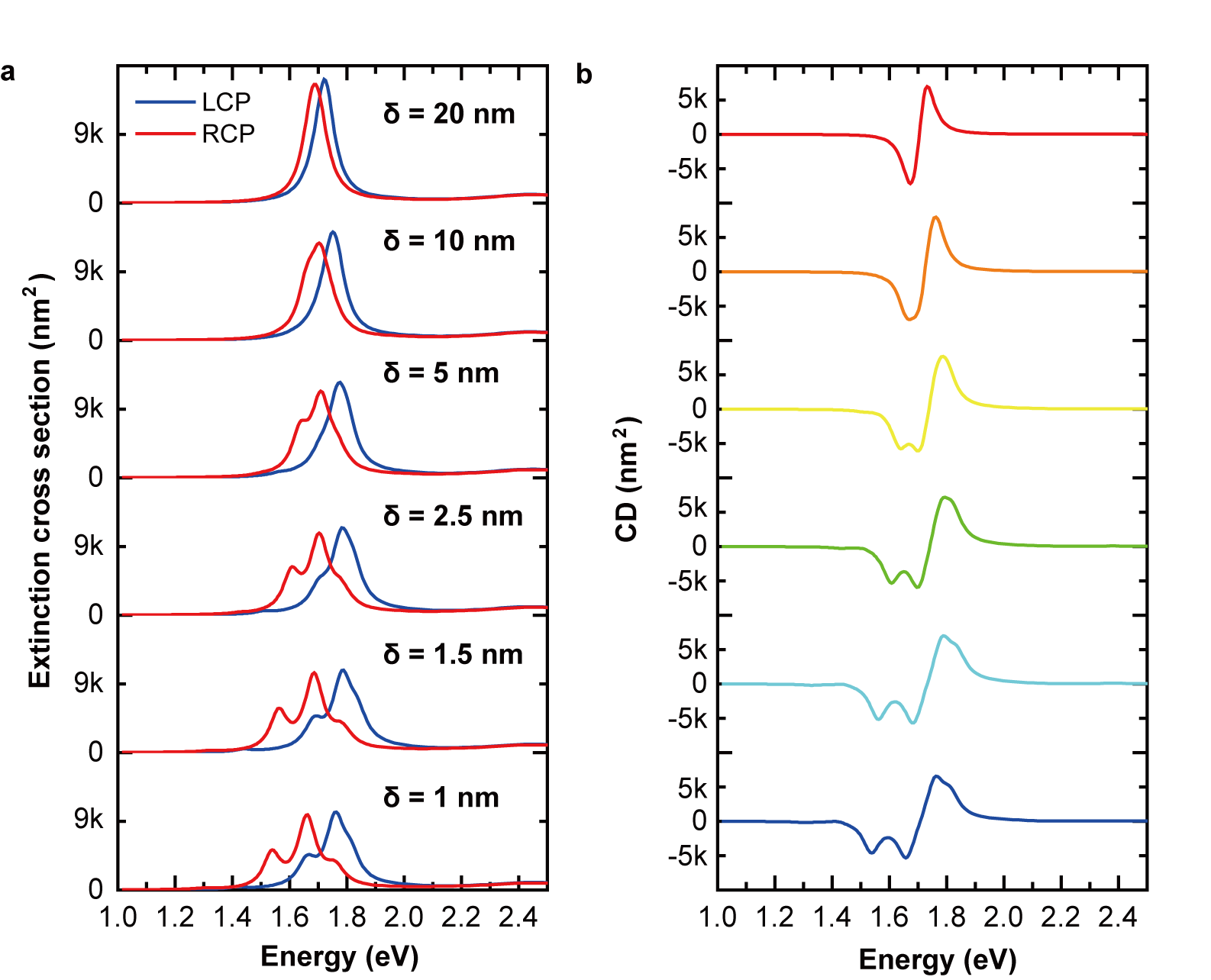}
    \caption{Investigation of the gap-dependent chiroptical responses by FEM-simulation. (a) Gap distance-dependent extinction spectra of the helical structures (the same configuration as FIG.~\ref{fig:Fig4_helix3_2}) excited by LCP (blue line) and RCP (red line). The gap distance is increased from $1$ to $20~\mathrm{nm}$. (b) Corresponding CD spectra of the helical structures. The hybridized peaks almost disappear at the gap distance at $20~\mathrm{nm}$, which indicates that only the dipolar oscillation of individual nanorods is present in the helical structures at large gap distance.}
    \label{fig:Fig5_gap}
\end{figure}

A key feature of the end-to-end assembled nanorod helical structure is the distance-dependent coupling between adjacent nanorods. To investigate the effect of the coupling strength on the resonances, we present in FIG.~\ref{fig:Fig5_gap} the gap distance-dependent extinction spectra of the helical structures consisting of two successive unit cells. For a small gap distance of $1~\mathrm{nm}$, the nanorods are coupled strongly to each other, giving rise to strongly hybridized resonance modes, where dipolar or higher-ordered oscillations in individual nanorods are present. With the increase of the gap distance, the coupling and the multiple hybridized resonance peaks tend to merge into a single one (FIG.~\ref{fig:Fig5_gap}). At the largest gap distance of $20~\mathrm{nm}$, a single resonance peak with no trace of hybridization is observable, where only dipolar oscillation in individual nanorods is present. This scenario is analogous to the nanorod helical structures with the lateral arrangement, and the chiroptical response is solely determined by the orientations of the nanorods in the helical structure\cite{lan2014nanorod}. 

In the following discussion, we employ only the classical electromagnetic model and set the minimum gap distance as $1~\mathrm{nm}$. In reality,the gap between adjacent nanoparticles in a typical DNA origami-assembled nanoparticle superstructures is hard to reach the sub-nanometer region, because of the spatial barrier provided by the DNA origami template as well as the surface functionalization of the nanoparticles. However, for the structure with the particle separations within a few angstroms, one needs to be aware of the quantum-mechanical effects, such as inter-particle electron tunneling, which induces a damping effect on the coupling resonance. In this case, a quantum-corrected model is needed to investigate such coupling damping due to the quantum electron tunneling between adjacent particles\cite{tserkezis2018circular,esteban2012bridging}. Here, we will not go deep into this model as the effect is negligible in our superstructure under investigation. 

\begin{figure}[tpb]
    \centering
    \includegraphics[width=1.0\linewidth]{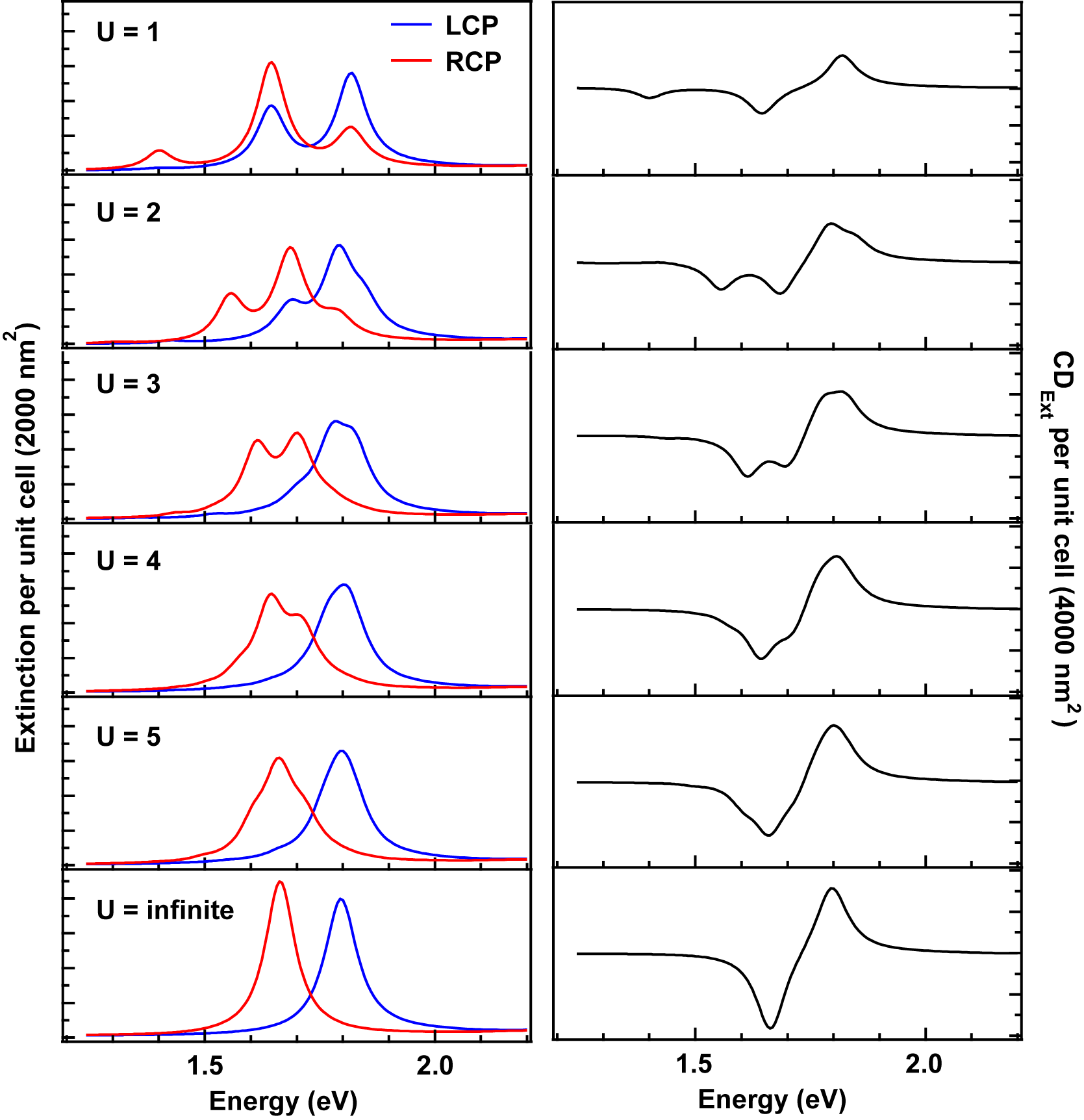}
    \caption{Investigation of the chiroptical responses on the number of the unit cell by FEM-simulation. Left panel: Evolution of the LCP and RCP extinction spectra of the helical structures when the unit cell is repeated along the helical axis and forms a superstructure. The number of the unit cell is 1, 2, 3, 4, 5, and infinity from the top to the bottom. In this case, one of the unit cell consisting of three nanorods (the same configuration as FIG.~\ref{fig:helix_3_mode}). Right panel: Evolution of corresponding CD spectra. Please noted that the hybridized peaks tend to merge into a single peak when the number of the unit cell continues to increase. For helical structure contains five cells, the hybridized peaks almost disappear. For an infinite helical superstructure, a stand-alone single peak can be observed at $1.66~(1.80)~\mathrm{eV}$ for RCP (LCP).}
    \label{fig:Fig6_inf}
\end{figure}

In the previous sections, we have investigated the nanorod helical structures consisting of only one or two unit cells. When the number of the unit cell, $U$, is increased, one may naturally expect richer features of the optical response from the structure, as the complexity of the structure and the opportunities of the inter-particle coupling are increased. However, the system will also become more symmetric, and hence strong inter-particle coupling leads to a collective mode dominating in the helical structure. Competing with collective resonance, those of the individuals become less significant. In order to clearly depict this evolution of the resonance, the unit cell consisting of three nanorods is repeated along the helical axis and a superstructure is formed. FIG.~\ref{fig:Fig6_inf} shows the FEM-simulated results for such an evolution from $U = 1$ to infinity. In a single-unit cell, the resonance of the nanorod splits into three distinctive modes due to strong hybridization. The mode at the highest energy is dominated by LCP, while the other two modes at lower energies are dominated by RCP. When the helical structure contains two successive unit cells, six hybridized peaks are observable in the spectrum and they tend to merge into a single peak at energy depending on LCP and RCP. Further merging of the peaks is observed when the number of the unit cell continues to increase. When the helical structure contains five cells, the peaks are merged together so that the shoulders of the peak are not so obvious. Finally, for an infinite helical superstructure, a stand-alone single peak can be observed at $1.66~(1.80)~\mathrm{eV}$ for RCP (LCP), indicating the pure character of RCP (LCP).

The evolution of the modes in the helical structure can be explained qualitatively. Strong hybridization dominates in the single-unit cell, which gives rise to well-defined hybridized modes at discrete energies. In a helical structure with a few unit cells, a collective mode driven by nanorods in the middle part of the chain starts to appear due to the periodic feature of the chain while the nanorods near the ends of the chain still exhibit individual characters. The hybridization is weakened as more nanorods participate in the collective mode, which results in frequency detuning by off-resonance oscillation as well as smaller gaps of the hybridized energies. When the number of the unit cells is further increased, the off-resonance effect becomes more significant, resulting in the merging of the resonance peaks and appearance of the collective mode. Finally, for an infinitely long helical superstructure, the collective mode dominates and surpasses all the hybridized modes with individual features, resulting in a single resonance peak in the spectra. As the collective mode is determined by both the symmetry of the helical structure and the circular polarization of the incident light, it responses exclusively to LCP or RCP at different frequencies, and hence this well-defined polarization dependence leads to the characteristic dip-peak signature of the CD signals.

\section{Generlized theory and discussion}
\begin{figure}[tpb] 
    \centering
    \includegraphics[width=0.8\linewidth]{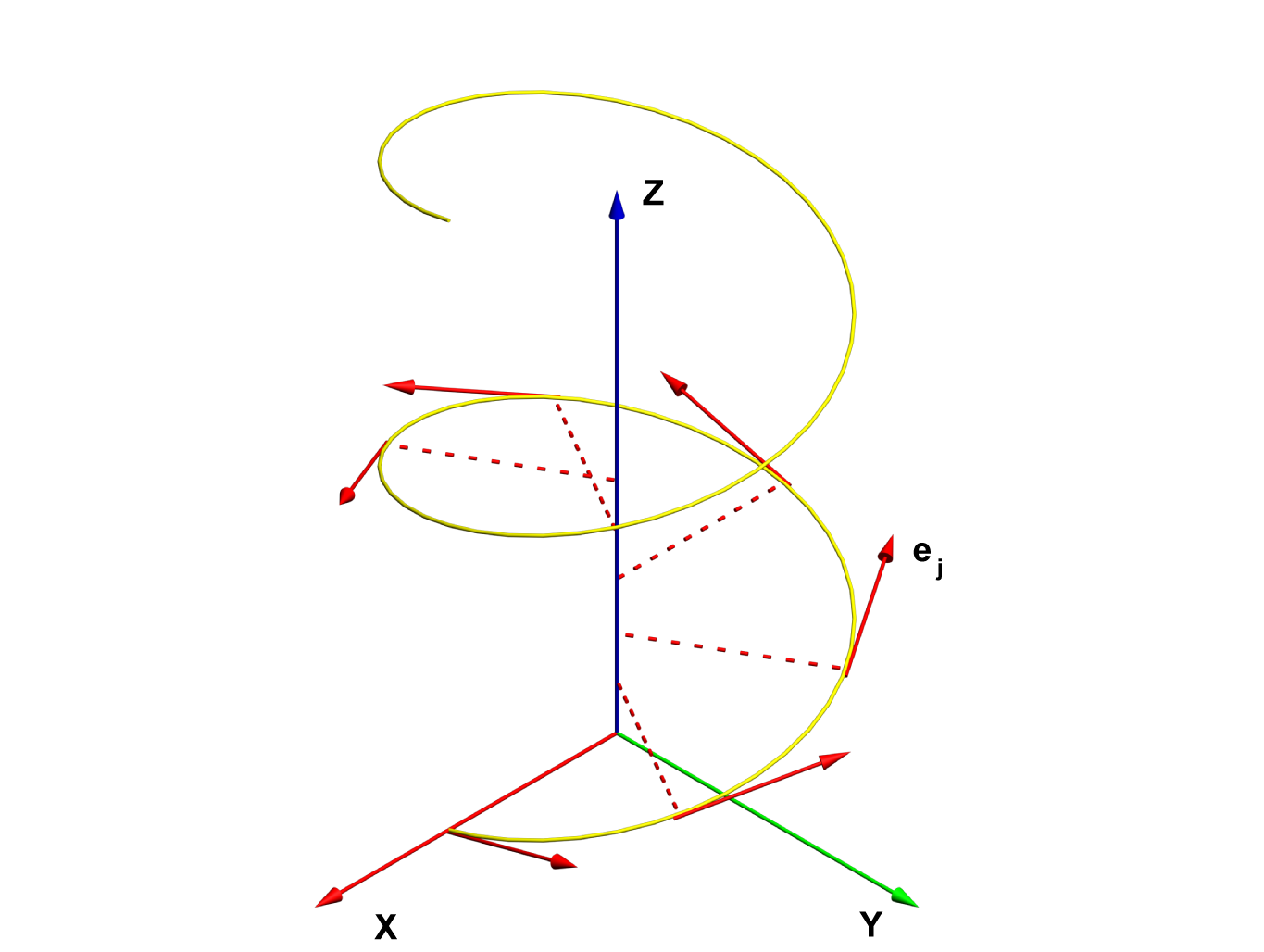}
    \caption{Model of helical dipoles used in the CDA calculation. Each nanorod in the helical structure is modeled as a dipole located on the helix and orients along its tangent with a certain gap.}
    \label{fig:dipole_x}
\end{figure}

According to the aforementioned numerical simulations, we learn that the helical superstructure possesses a collective resonance at a specific energy, which depending on the LCP or RCP of the incident light. However, the origin of the selection rule of the collective resonance mode is still elusive.   To have a generalized theory to describe the resonance mode in the helical superstructure, we propose an analytical model by employing the CDA approach to theoretically describe the collective resonance and its interactions with the CPL\cite{yurkin2016corrigendum}.

For sake of simplicity, in the following equations, the bold letters denote vectors, Dirac-notation represents the product of polarizations and dot-notation represents the product of the phase. In our CDA calculation, each nanorod is small enough to be modeled as a dipole and ignore the lateral coupling between adjacent nanorods. The gold nanorods in the helical structure are located at:
\begin{equation}
    \mathbf{r}_j=R\cos\varphi_j\mathbf{e}_x + R\sin\varphi_j\mathbf{e}_y +R\varphi_j\tan\theta\mathbf{e}_z,
    \label{dipoleR}
\end{equation}
where j donote the $j-$th nanorod and it orients along the tangent of the helix(FIG.~\ref{fig:dipole_x}),
\begin{equation}
    \mathbf{e}_j=-\cos\theta\sin\varphi_j\mathbf{e}_x+\cos\theta\cos\varphi_j\mathbf{e}_y+\sin\theta\mathbf{e}_z.
    \label{dipole_v}
\end{equation}
Only the longitudinal oscillation is considered in our theory, and hence the dipole and polarizability tensor of each nanorod can be expressed as:
\begin{equation}
\begin{split}
    &\mathbf{p}_j = p_j\mathbf{e}, \\
    &\mathbf{\alpha}_j = \alpha_0\mathbf{e}_j\mathbf{e}_j.
    \label{eq:dipole_v}
\end{split}
\end{equation}
Where $p_j$ is the strength of the $j$-th dipole and $\alpha_0$ is the major polarizability which can be obtained by the Rayleigh-Gans Approximation. Considering a one-dimensional periodic array consisting of n dipoles spaced by a displacement of $\mathbf{D}$, one can express the basic unit state of the dipole polarization as:
\begin{equation}
    P=\left[\mathbf{p}_1~\mathbf{p}_2~\dots~\mathbf{p}_n\right]^T,
    \label{eq:dipole_array}
\end{equation}
and it satisfies the self-consistent coupled equation\cite{auguie2008collective,de2007colloquium,draine2008discrete}:
\begin{equation}
    \alpha^{-1}P=E+S(\mathbf{k}\cdot\mathbf{D})P.
    \label{eq:couple_P}
\end{equation}
Here $\alpha$ is a diagonal matrix containing the electric polarizability of each dipole:
\begin{equation}
    \alpha = \left[
    \begin{array}{cccc}
	\mathbf{\alpha}_1 &0 &\dots &0\\
	    0 &\mathbf{\alpha}_2 &\dots &0\\
	    \vdots &\vdots &\ddots &\vdots\\
	    0 &0 &\dots &\mathbf{\alpha}_n
    \end{array}
   \right],
    \label{eq_alpha_array}
\end{equation}
and
\begin{equation}
    E=\left[\mathbf{E}_1~\mathbf{E}_2~\dots~\mathbf{E}_n\right]^T
    \label{eq_dipole_E}
\end{equation}
represents external electric fields acting on the dipoles.
\begin{equation}
    S(\mathbf{k}\cdot\mathbf{D})=\sum_{N=-\infty}^{N=+\infty}G(N\mathbf{D})\exp(iN\mathbf{k}\cdot\mathbf{D}),
    \label{eq:matrix_sum}
\end{equation}
is the sum of the interactions over all the dipoles both on the target and replica units, where $\exp(iN\mathbf{k}\cdot\mathbf{D})$ denotes the Bloch's phase shift of the $N$-th replica unit and the matrix $G(N\cdot\mathbf{D})$:
\begin{widetext}
\begin{equation}
    G(N\mathbf{D}) = \left[
    \begin{array}{cccc}
	\mathbf{G}(\mathbf{r}_1,\mathbf{r}_1+N\mathbf{D}) &\mathbf{G}(\mathbf{r}_1,\mathbf{r}_2+N\mathbf{D}) &\dots &\mathbf{G}(\mathbf{r}_1,\mathbf{r}_n+N\mathbf{D})\\
	\mathbf{G}(\mathbf{r}_2,\mathbf{r}_1+N\mathbf{D}) &\mathbf{G}(\mathbf{r}_2,\mathbf{r}_2+N\mathbf{D}) &\dots &\mathbf{G}(\mathbf{r}_2,\mathbf{r}_n+N\mathbf{D})\\
	    \vdots &\vdots &\ddots &\vdots\\
	\mathbf{G}(\mathbf{r}_n,\mathbf{r}_1+N\mathbf{D}) &\mathbf{G}(\mathbf{r}_n,\mathbf{r}_2+N\mathbf{D}) &\dots &\mathbf{G}(\mathbf{r}_n,\mathbf{r}_n+N\mathbf{D})\\
    \end{array}
   \right],
    \label{eq_G_array}
\end{equation}
\end{widetext}
contains all the Green tensors between the target and the $N$-th replica unit, separated by a displacement of $N\mathbf{D}$. By defining effective polarizability as:
\begin{equation}
    \alpha_\mathrm{eff}^{-1}=\alpha^{-1}-S(\mathbf{k}\cdot\mathbf{D}),
    \label{eq:alpha_eff}
\end{equation}
Eq.~\eqref{eq:couple_P} can be written as:
\begin{equation}
    \alpha_\mathrm{eff}^{-1}P=E.
    \label{eq:eff_couple}
\end{equation}
Once this self-consistent equation is solved, we can calculate the cross-sections of extinction, absorption, and scattering of the system. On the basis of Eq.~\eqref{eq:eff_couple}, the dispersion for the collective modes can be represented by\cite{weber2004propagation,koenderink2006complex,lunnemann2014dispersion}
\begin{equation}
    \mathrm{Det}[\alpha_\mathrm{eff}^{-1}]=0.
    \label{eq:eff_dispersion}
\end{equation}
\begin{figure}[tpb]
    \centering
    \includegraphics[width=1.0\linewidth]{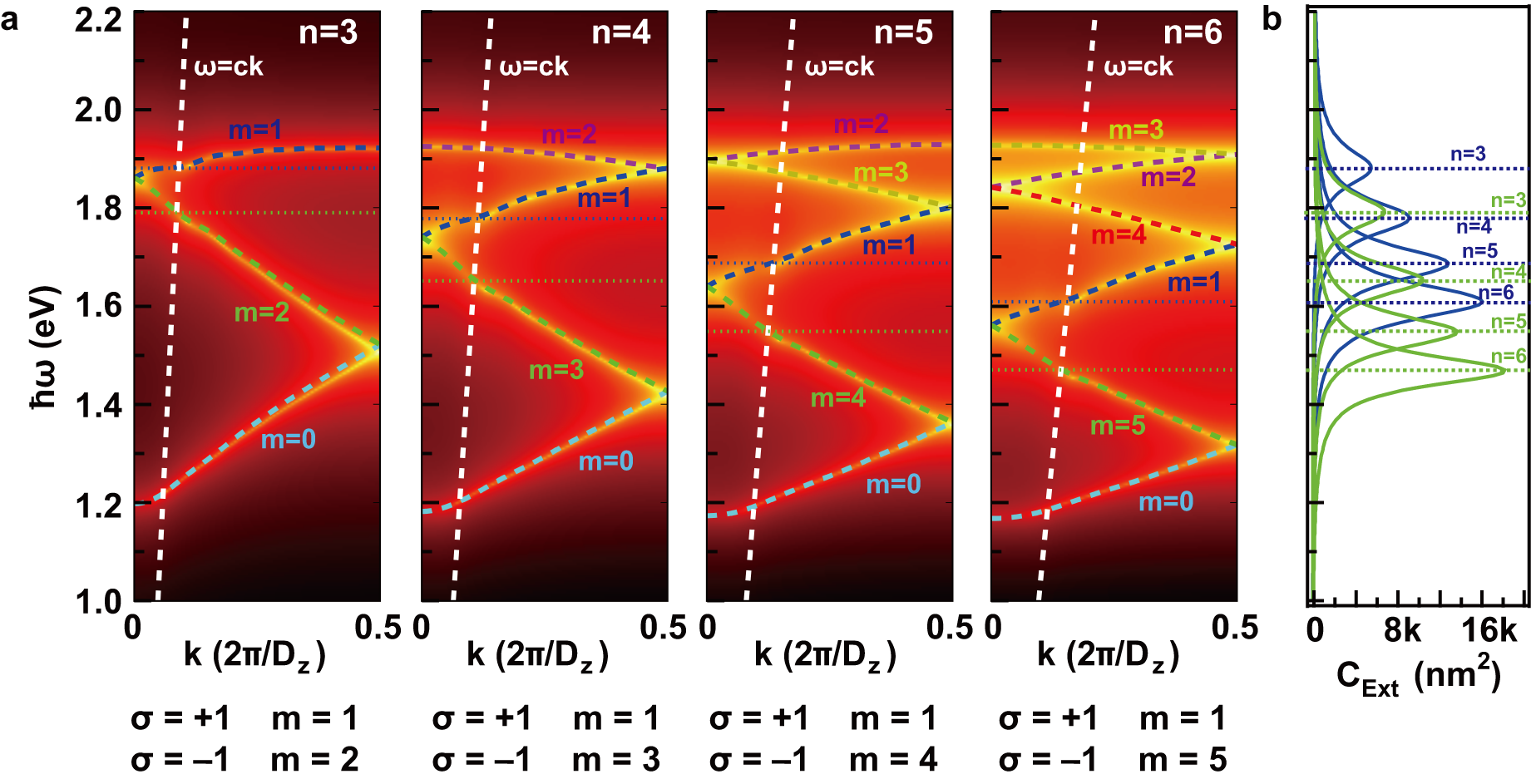}
    \caption{CDA calculation of the of collective resonance modes. In our calculations, each gold nanorod is approximated as a prolate ellipsoid. the major and minor radii of the ellipsoid were set as $6.1$ and $22.265~\mathrm{nm}$ respectively. The vertical distance between the nearest ellipsoid in all the structures is set at 30 nm.  (a) Dispersion contours of collective resonance modes in infinite helical superstructures. The investigated helical superstructures with $n = 3,~ 4, ~5$, and $6$ are presented from left to right. The dispersion curves with azimuthal index $m$ are plotted as the dashed lines in different colors. The light cone is shown as the white dashed line. The selection rule determines the active mode under the excitation of LCP or RCP. (b) Corresponding extinction spectra under the excitation of LCP (blue) and RCP (green).}
    \label{fig:CDA_dispersion}
\end{figure}

To obtain a clear physical picture of collective resonance modes, we calculate their dispersions as well as thier interactions with CPL. Four infinite helical superstructures with $n = 3,~4,~5$, and 6 are investigated. $\ln[|\mathrm{Det}(\alpha_\mathrm{eff}^{-1})|^{-1}]$ is evaluated and contoured as a function of  in FIG.~\ref{fig:CDA_dispersion}a. For simplicity, the loss in the nanorods is neglected in the CDA calculation, and the vertical distance between the nearest dipoles in all the structures is set at $30~\mathrm{nm}$. Since the structure is a one-dimensional periodic structure, the summation is able to converge, we truncate the infinite sum of  Eq.~\eqref{eq:matrix_sum} by setting the interval of $N$ with $N\in[-30,30]$. The bright stripes represent the dispersions of the resonance modes. In total $n$ collective resonance modes are excited for the helical superstructure formed by the unit cell consisting of n nanorods. The dispersions of the resonance modes are separated from each other in frequency but intersect with each other at the edges of $k_z$. To be more precise, the dispersions of the resonance modes are separated in half of the Brillouin zone. The spectra of extinction cross-section are calculated and plotted for LCP and RCP on the basis of the CDA model (FIG.~\ref{fig:CDA_dispersion}b).

Next, we continue our discussion with the mode analysis for the helical superstructure on the basis of symmetry. Besides the translational symmetry, the superstructure possesses the symmetry element related to the helical structure, and hence the collective resonance mode should represent the additional symmetry features other than the Bloch's phase shift from the translational symmetry. For the helical superstructure formed by the unit cell consisting of $n$ nanorods, the symmetry can be viewed as a combination of an $n$-fold rotation $R_{2\pi/n}$ and a translation $T_{D_z/n}$ along the rotation axis. In the meanwhile, the periodic symmetry requires the relation $(T_{D_z /n}\cdot R_{2\pi /n})^n=T_{D_z}$. Applying these operations on any scale field $\Psi$, 
\begin{equation}
\begin{split}
    (T_{D_z /n}\cdot R_{2\pi/n})^n\Psi&=T_{D_z}\Psi=\exp(ik_mD_z)\Psi\\
    \to(T_{D_z /n}\cdot R_{2\pi/n})^n\Psi&=\exp(ik_mD_z+i2m\pi)\Psi
\end{split}
    \label{eq:Group_syms}
\end{equation}
where $0\le k_m\le 2\pi/D_z$, and $m=0,~1,~\dots,~n-1$. Obviously, all the irreducible representations of the helix group are one-dimensional. From Eq.~\eqref{eq:Group_syms}, one can easily find that the l-th character of the $m$-th irreducible representation can be expressed as $\exp[il(k_mD_z/n+2m\pi /n)]$. By using the projection theorem of the group theory\cite{cotton2003chemical}, we get $n$ eigenmodes for the helical chains of the dipoles, and the $m$-th mode labeled by azimuthal index m can be written as:
\begin{equation}
    p_j^m\mathbf{e}_j=p_0^m\exp[i(k_pz_j+m\varphi_j)]\mathbf{e}_j.
    \label{eq:collective_mode}
\end{equation}
Here we choose coordinates $z_0=0$ and $\varphi_0=0$, and then we have $z_j=jD_z/n$ and $\varphi_n=j 2\pi /n$ for the $j$-th particle.

Now we consider the collective oscillation excited by the CPL propagating along the z-direction. In this case, the dipole polarization can be expressed as the linear superposition of the helical eigenmodes:
\begin{equation}
    p_j = \sum_{m} C_mp_j^m,
    \label{eq:eigenmode}
\end{equation}
where $C_m$ is the expansion pre-factor. Employing the CDA approach for each dipole leads to:
\begin{equation}
    \frac{p_i}{\alpha_0}-\sum_{j\ne i}\Braket{\mathbf{e}_j|\mathbf{G}(\mathbf{r}_i,\mathbf{r}_j)|\mathbf{e}_j}p_j = \Braket{\mathbf{e}_i|\mathbf{E}_i}.
    \label{eq: CDA_eigenmode}
\end{equation}
We then have
\begin{equation} 
    \sum_m C_m p_0^m\exp[i(k_pz_i+m\varphi_i)]\frac{1}{\alpha_\mathrm{eff}^m}=\Braket{\mathbf{e}_i|\mathbf{E}_i},
    \label{eq:eigenmode_couple}
\end{equation}
where
\begin{small}
\begin{equation}
    \frac{1}{\alpha_\mathrm{eff}^m}=\frac{1}{\alpha_0}-\sum_{k\ne0}\Braket{\mathbf{e}_i|\mathbf{G}(\mathbf{r}_i,\mathbf{r}_{i+k})|\mathbf{e}_{i+k}}\exp[i(k_p z_k+m\varphi_k)]
    \label{eq:eigenmode_eff}
\end{equation}
\end{small}
is independent of the particle label due to the helical symmetry, and it can be viewed as the effective polarizability of each dipole for the $m$-th helical mode.

So far, we have described the collective resonance eigenmodes of the helical superstructure. Based on the derivation, one can identify these modes by solving the equation $(\alpha_\mathrm{eff}^m)^{-1}=0$ numerically. They are plotted in FIG.~\ref{fig:CDA_dispersion}a as dashed lines representing the dispersion relations for modes with a different azimuthal index of $m$. As expected, a good agreement is found between the dispersion curves of $m$-th resonance mode and the bright stripes in the contour. The interactions of these modes with the incident light can be represented as the intersection between the light cone and the dispersion curves. The light cone intersects with all the dispersion curves, suggesting the possibility of exciting all the modes by the planar light waves. However, considering certain handedness of the incident light, LCP or RCP, we have to include it into our calculation.

The incident field of the CPL at the location of i-th dipole can be expressed as
\begin{equation} 
    \mathbf{E}_i=\mathbf{e}_\sigma\exp(ikz_i),
    \label{eq:Efield}
\end{equation}
where $\mathbf{e}_\sigma=\frac{1}{\sqrt{2}} (\mathbf{e}_x+i\sigma\mathbf{e}_y)$ are circular polarization vectors, with $\sigma=\pm 1$ representing the LCP or RCP, respectively. Using these vectors, the direction of each dipole polarization can be written as
\begin{equation} 
    \mathbf{e}_i=-i\frac{\cos\theta}{\sqrt{2}}\sum_{\sigma=\pm}\sigma\exp(-i\sigma\varphi_i)\mathbf{e}_{\sigma}+\sin\theta\mathbf{e}_z,
    \label{eq:CPL_direction}
\end{equation}
and hence we have
\begin{equation} 
    \Braket{\mathbf{e}_i|\mathbf{E}_i}=i\sigma \frac{\cos\theta}{\sqrt{2}}\exp(ikz_i+i\sigma\varphi_i).
    \label{eq:Efield_pi}
\end{equation}
By substitute Eq.~\eqref{eq:Efield_pi} into Eq.~\eqref{eq:eigenmode_couple}, we have
\begin{small}
\begin{equation} 
    \sum_m C_mp_0^m\exp[i(k_pz_i+m\varphi_i)]\frac{1}{\alpha_\mathrm{eff}^m}=i\sigma \frac{\cos\theta}{\sqrt{2}}\exp[i(kz_i+i\sigma\varphi_i)].
    \label{eq:eigenmode_dispersion}
\end{equation}
\end{small}
Eq.~\eqref{eq:eigenmode_dispersion} must hold for the $i$-th dipole in the infinite array, which requires the phase of both sides of the equation should match. Therefore for $C_m \ne0$, we have $k_p=k$ and a selection rule of excitation ($\sigma=1$, $m=1$; $\sigma=-1$, $m=n-1$). It means that only one mode can be excited by the CPL among the $n$ collective helical modes, depending on the handedness of the incident CPL. Actually, this requirement by the phase match determines that only one mode can be effectively excited and hence be active, which is confirmed by the single peak in the spectrum from either numerical simulation (FIG.~\ref{fig:Fig6_inf}) or CDA calculations (FIG.~\ref{fig:CDA_dispersion}b). Our result shows that the resonance energies of the peaks fit with the intersections between the light cone and the dispersion curve labeled by $m=1$ for LCP and $m=n-1$ for RCP (FIG.~\ref{fig:CDA_dispersion}). With the dispersion relation and the selection rule, one can simply predict the resonance energy of the collective mode of a well-defined superstructure without calculating the whole spectrum.

From the above discussions, we find that the number of existing modes is determined by the order of the discrete rotational symmetry. Eq. (15) still holds true for the helical structure when the order $n$ of the symmetry increases to the infinite, which means there exist infinite modes in the continuous helical structure, such as the helicoidal waveguides response in THz region\cite{fernandez-dominguez2008terahertz,ruting2012subwavelength}. However, the full-wave analysis is needed for such continuous structures as the coupled dipole approximations are no longer valid. On the contrary, discrete gold nanorod helical superstructures in a general case can be well described by our coupled dipole modes. Specifically, each mode in the superstructures is well separated and the collective modes are from the coupling between the discrete nanorods.

At last, we should mention that our model is not only restricted to the simple case in this study, where the circular polarization light is interacting with the end-to-end gold nanorod helix structure. Expanding to more general cases is straightforward and we will not go into details in this paper. For instance, one can simply replace the circular polarized incident field in Eq~\eqref{eq:Efield} with optical helix beam to study the orbital angular momentum interaction of light with our gold nanorod end-to-end helical structures, or modify the related positions in Eq~\eqref{dipoleR} or orientations in Eq~\eqref{dipole_v} of each nanoparticles to investigate the optical response of more complicated structure. 

\section{Summary}
In conclusion, we have provided a detailed theoretical description for the collective resonance in a typical helical gold nanorod superstructures. Such helical gold nanorod superstructures may be realized in the future by using the methods of self-assembly, such as DNA-origami directed assembly.  Numerical simulations reveal that the significantly well separated strong coupling in different resonance modes. Such resonance modes are in favor of certain circular polarizations, depending on their charge profiles on the nodes, thus resulting in a strong chiroptical response. With the increase of the helical periods,  the hybridized resonance peaks gradually evolve and merge into one, corresponding to a collective resonance mode whose energy is dependent on LCP and RCP. Therefore, we proposed an analytical model for the collective resonance mode of the infinite helical structure based on the CDA method. Our results show that there exist $n$ collective modes of $n$ fold-helical structure and each mode has the azimuthal angle-dependent phase in the form of $\exp(im\varphi_i)$ , labeled by the azimuthal index of $m$. To predict the collective resonance modes of helical gold nanorod superstructures, we reported a selection rule for the excitations of the collective resonance of the helical superstructure, which may promote the designs and applications of such helical superstructures in future.

\section*{Acknowledgment}
This work was financially supported by the National Natural Science Foundation of China (grants 21271181 and 21473240).

\section*{Appendix: Fem simulation and CDA model} 
\subsection{Finite element method (FEM)}
In this work, Comsol Multiphysics was employed to investigate the chiroptical response of the end-to-end assembled nanorod helical structures. In these simulations, the nanorod was modeled as a cylinder with two hemispheres at both ends, and the diameter and length were set at $12$ and $40~\mathrm{nm}$ respectively. The refractive index of the ambient medium was set at $1.33$.

\subsection{Coupled dipole approximation (CDA)} 

In the CDA model, each gold nanorod is approximated as a prolate ellipsoid. For simplicity, we only consider the longitudinal oscillation of the ellipsoid, which can be represented by the longitudinal polarizability $\alpha_0$. By using the Rayleigh-Gans Approximation, $\alpha_0$ polarizability can be obtained as\cite{bohm1957causality}
\begin{equation}
    \alpha_0=\frac{3V}{4\pi}\frac{\varepsilon-\varepsilon_m}{\varepsilon_m+(\varepsilon-\varepsilon_m)L},
    \label{eq:APPD_alpha}
\end{equation}
where $V$ is the volume of the nanoparticle,$\varepsilon_m$ is the dielectric constant of the surrounding medium, and $\varepsilon$ is the dielectric function of gold. The depolarization factor is defined as\cite{bohm1957causality}
\begin{equation} 
    L=\frac{1-e^2}{e^2}\left[\frac{1}{2e}\ln(\frac{1+e}{1-e})-1\right],
    \label{eq:APPD_L}
\end{equation}
where the eccentricity $e$ is defined as $e=\sqrt{1-b^2 /a^2} $ , $a$ and $b$ are the major and minor radii of the ellipsoid, respectively. In the CDA calculation, we set $b=6.1~\mathrm{nm}$, $a=3.65b$ to reproduce the spectrum of the single nanorod calculated by the finite element method. The Green tensor in Eq.~\eqref{eq_G_array} can be expressed as\cite{draine1994discrete}
\begin{widetext}
\begin{equation} 
    \mathbf{G}(\mathbf{r}_j,\mathbf{r}_i) = \frac{\exp(ikr_{ji})}{r_{ji}^3}\left[\frac{1-ikr_{ji}}{r_{ji}^2}
    \times (3\mathbf{r}_{ji}\mathbf{r}_{ji}-r_{ji}^2)-k^2\mathbf{r}_{ji}\times\mathbf{r}_{ji}\right],
    \label{eq:APPD_Green}
\end{equation}
\end{widetext}
where $\mathbf{r}_{ji}=\mathbf{r}_j-\mathbf{r}_i$, $r_{ji}=|\mathbf{r}_j-\mathbf{r}_i |$.

In the calculation of the summing parameter $S$ for the one-dimensional periodic helical structure, we truncate the infinite sum of  Eq.~\eqref{eq:matrix_sum} by setting the interval of $N$ with $N \in[-30,30]$. By solving these equations for the unknown polarizations $\mathbf{p}_j$ with tabled dielectric function\cite{johnson1972optical}, the extinction cross-section and absorption cross-section can be calculated as\cite{draine1994discrete}:
\begin{equation} 
    C_\mathrm{ext}=\frac{4\pi k}{|\mathbf{E}_\mathrm{inc}|^2}\sum_{j}\Im(\mathbf{E}_{\mathrm{inc},j}^{*}\cdot\mathbf{p}_j),
    \label{eq:APPD_Cext}
\end{equation}
and 
\begin{equation} 
    C_\mathrm{abs}=\frac{4\pi k}{|\mathbf{E}_\mathrm{inc}|^2}\sum_{j}\lbrace\Im[\mathbf{p}_j\cdot(\alpha_j^{-1})^*\cdot\mathbf{p}_j^*]-\frac{2}{3}k^3\mathbf{p}_j\cdot\mathbf{p}_j^*\rbrace.
    \label{eq:APPD_Cabs}
\end{equation}
%\bibliography{Reference}

\end{document}